\documentclass[12pt,superscriptaddress,showpacs,showkeys,twocolumn,prb,aps,reprint,a4paper,floatfix]{revtex4-1}

\usepackage{graphicx}
\usepackage{amsmath, amsthm, amssymb, amsfonts}
\usepackage{float}

\usepackage{natbib}
\usepackage{physics}

\usepackage[para,online,flushleft]{threeparttable} 

\usepackage{array}
\usepackage{ragged2e} 

\makeatletter
\newcommand*{\citenst}[2][]{%
  \begingroup
  \let\NAT@mbox=\mbox
  \let\@cite\NAT@citenum
  \let\NAT@space\NAT@spacechar
  \let\NAT@super@kern\relax
  \renewcommand\NAT@open{[}%
  \renewcommand\NAT@close{]}%
  \citet[#1]{#2}%
  \endgroup
}
\makeatother

\makeatletter
\newcommand*{\citenumns}[2][]{%
  \begingroup
  \let\NAT@mbox=\mbox
  \let\@cite\NAT@citenum
  \let\NAT@space\NAT@spacechar
  \let\NAT@super@kern\relax
  \renewcommand\NAT@open{[}
  \renewcommand\NAT@close{]}%
  \cite[#1]{#2}
  \endgroup
}
\makeatother
\usepackage[caption=false]{subfig}

\begin{document}
\title{Loss and Saturation in Superconducting Travelling-Wave Parametric Amplifiers}
\author{Songyuan Zhao}
\email{sz311@cam.ac.uk}
\author{S. Withington}
\author{D. J. Goldie}
\author{C. N. Thomas}
\date{\today}

\affiliation{Cavendish Laboratory, JJ Thomson Avenue, Cambridge CB3 OHE, United Kingdom.}

\begin{abstract}
\noindent We have developed a coupled-mode analysis framework for superconducting travelling-wave parametric amplifiers using the full Telegrapher's equations to incorporate loss-related behaviour. Our model provides an explanation of previous experimental observations regarding loss in amplifiers, advantages of concatenating amplifiers to achieve high gains, and signal gain saturation. This work can be used to guide the design of amplifiers in terms of the choice of material systems, transmission line geometry, operating conditions, and pump strength.
\end{abstract}


\maketitle

\section{Introduction}
Recent decades have seen rapid developments in the fields of superconducting quantum information processing and cryogenic photon detectors \citenumns{Tholen_2009,Clerk_2010}. Integral to reading out the thin-film superconducting devices are high gain, broad-band, low-noise amplifiers. In many cases, these amplifiers are based on High Electron Mobility Transistors (HEMTs), but this imposes restrictions on the sensitivities of readout systems at several times the Standard Quantum Limit \citenumns{Eom_2012,McCulloch_2017}. Now, however, superconducting amplifiers are being studied as alternatives to transistor-based amplifiers, with the aim of achieving the quantum limit, and indeed pushing below the quantum limit using quadrature squeezing techniques. Josephson junction amplifiers have demonstrated quantum-limited noise performance, and have been integrated into various qubit and optomechanics systems \citenumns{Regal_2011,Shankar_2013}, but they tend to be designed around resonant structures, and therefore have narrow bandwidths, of the order of a few megahertz \citenumns{Vijay_2011}. In addition, as these devices typically have low saturation powers \citenumns{Castellanos_2010}, they are unsuitable for applications such Kinetic Inductance Detector (KID) readout \citenumns{Bockstiegel_2014}. Applications that require large bandwidths, such as qubit and detector readout and multiplexing \citenumns{Eom_2012}, have motivated experimental demonstrations of superconducting travelling-wave parametric amplifiers (TWPAs) which can have bandwidths of the order of a few gigahertz \citenumns{Eom_2012,Bockstiegel_2014,Adamyan_2016,Chaudhuri_2017}. In this case, the intrinsic non-linearity associated with the inductance of a thin-film superconducting transmission line acts as a non-linear medium, which under the appropriate conditions can lead to low-noise amplification.


Current analyses of TWPAs are based on the lossless Telegrapher's equations, which model signal propagation along a 1-dimensional transmission line as a pair of first-order differential equations \citenumns{Chaudhuri_2015,Shan_2016,Erickson_2017}. The solutions to the non-linear problem are then found by using either a coupled-mode solver, or a finite-difference-time-domain (FDTD) solver. Currently, coupled-mode solvers assume that the pump amplitude is not attenuated by the mixing process \citenumns{Eom_2012, Chaudhuri_2015}, which leads to predictions that the signal gain increases monotonically with amplifier length and is independent of the input signal amplitude. Existing FDTD solvers are computationally intensive \citenumns{Chaudhuri_2015,Shan_2016}, and the purely numerical approach means the results can be hard to understand intuitively.


The current body of literature on experimental realizations of TWPAs has drawn attention to a number of considerations that are yet to be described by existing analysis frameworks: The first is the effect of loss along the amplifier transmission line \citenumns{Eom_2012, Shan_2016}. The second is the saturation of the gain at high signal-to-pump ratios \citenumns{Eom_2012}. The third is the potential trade-off between a single long-length TWPA and concatenated multiple short-length TWPAs \citenumns{Chaudhuri_2017}.


In this study, we extend the analysis of TWPAs by using the full Telegrapher's equations, which include ohmic and dielectric loss, and thereby incorporate travelling-wave attenuation as well as parametric gain in our analysis. Further, we extend the coupled-mode solver to include higher-order mixing terms, enabling the analysis of TWPAs at high signal-to-pump ratios. We show that by including these extensions, our analysis framework is able to model practical considerations, such as loss, saturation, and optimum amplifier length. Moreover, this is done in a way that leads to conceptual insights that can guide the design of amplifiers when considering matters such as choice of superconducting material, operating temperature, and what line length and impedance to use.

\section{Analysis Framework}

The full Telegrapher's equations are used to model lossy transmission lines. The transmission line is described by an infinite series of differential elements of length $dz$. The pair of coupled differential equations are \citenumns{pozar2011microwave}
\begin{align}
  \pdv{}{z}V &= -L\pdv{}{t}I-RI \, \\
  \pdv{}{z}I &= -C\pdv{}{t}V-GV \, ,
\end{align}
where $V$ is the potential difference along the line, and $I$ is the current along the line. $L$, $C$, $R$, and $G$ are the  inductance, capacitance, resistance, and conductance per unit length, respectively, and $z$ and $t$ are the position and time coordinates.

The operation of a TWPA relies on the non-linearity of the kinetic inductance of superconductors to achieve wave mixing, and to allow energy transfer from the pump frequency to the signal frequency. The non-linear kinetic inductance is approximated by a quadratic expansion \citenumns{Jonas_review, Eom_2012}, and is introduced into the Telegrapher's equations by the substitution
\begin{equation}
  L = L_0[1+(I/I_*)^2] \, ,
\end{equation}
where $I_*$ is the scale of non-linearity. Once this has been done, the equations are non-linear, but can be linearised through certain approximations. $I_{*}$ is used to characterise the current at which the non-linear inductance becomes significant.

\subsection{Coupled-Mode Description}

Under the coupled-mode description, we express the current $I$ as a sum of all existing current modes $I_{\alpha}$
\begin{align}
  &I = \sum_{\alpha} I_{\alpha}\,,\label{eq:CM_sum}\\
  &I_{\alpha} =\frac{1}{2}A_{\alpha}(z)\operatorname{exp}(j\omega_{\alpha} t-\gamma_{\alpha} z)\,, \label{eq:CM_prop}\\
  &I_{\alpha}^* = I_{-\alpha}\,, \label{eq:CM_conj}
\end{align}
where $\omega_{\alpha}$ is the frequency of the current mode, $\gamma_{\alpha}$ is the complex propagation constant in the absence of inductive non-linearity, and $A_{\alpha}(z)$ is the position varying amplitude coefficient of the current mode. Here the exponential term contains the propagation behaviour of a linear lossy transmission line, and the $A_{\alpha}(z)$ term contains the effect of the non-linearity.

Under the coupled-mode description, the slowly-varying-envelope-approximation (SVEA) \citenumns{Boyd_2008, Chaudhuri_2015} is applied to give
\begin{align}
  \left|\pdv[2]{}{z}A_{\alpha}(z)\right|\ll \left|\gamma_{\alpha}\pdv{}{z}A_{\alpha}(z)\right| \,.
\end{align}
Then, by approximating $G\ll \omega_{\alpha} C$ for all $\omega_{\alpha}$ of interest, the Telegrapher's equations can be expanded to obtain
\begin{align}
-2\gamma_{\alpha}\frac{I_{\alpha}}{A_{\alpha}}\pdv{A_{\alpha}}{z} = \frac{L_0 C}{I_*^2}\pdv{}{t}\left(\sum_{i,j,k\rightarrow \alpha} I_i I_j\pdv{I_k}{t}\right)\,, \label{eq:Combination}
\end{align}
where the summation is applied over all frequency combinations that satisfy $\omega_i+\omega_j+\omega_k=\omega_{\alpha}$. When expanded, equation~(\ref{eq:Combination}) gives a set of coupled differential equations that needs to be solved to give $A_{\alpha}(z)$.

\subsection{Impedance Loading}

\begin{figure}[ht]
\includegraphics[width=8.6cm]{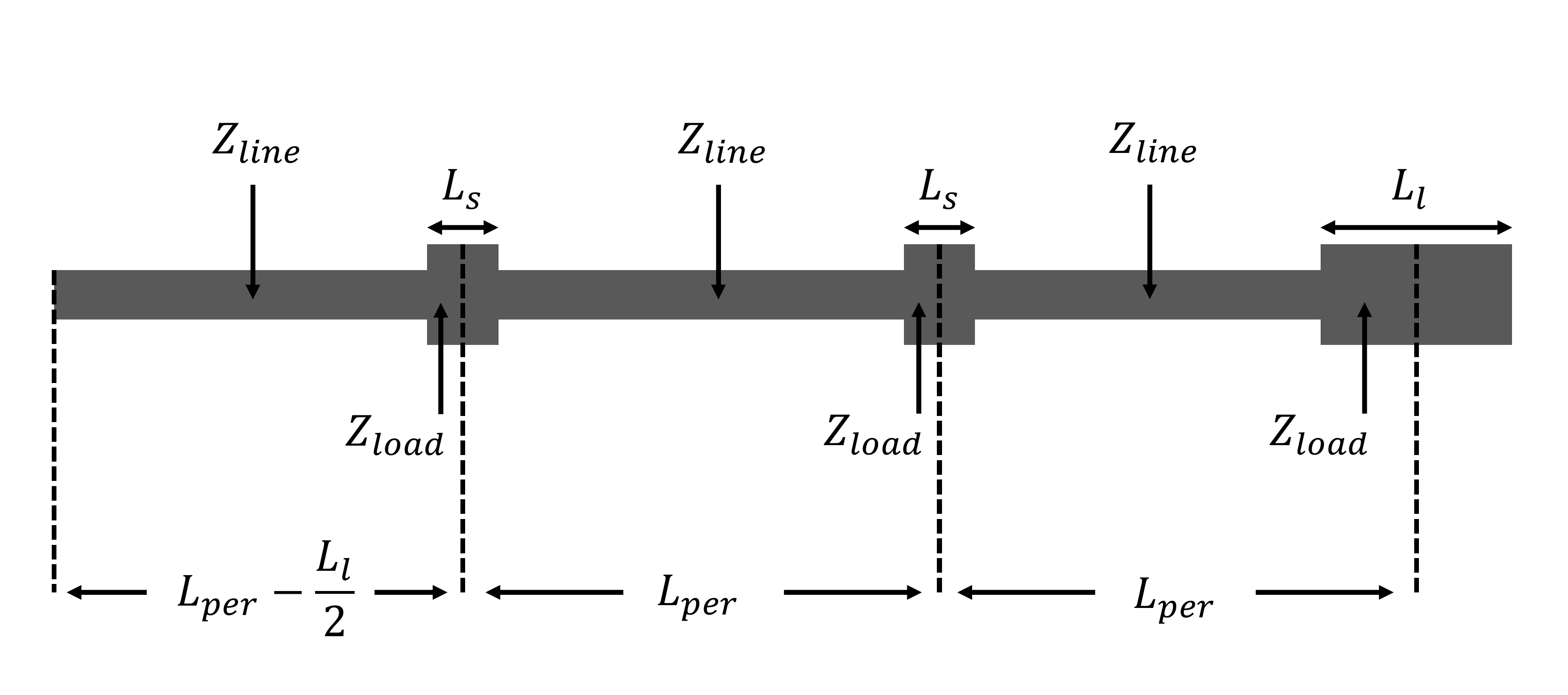}
\caption{\label{fig:Periodic_Loading} Schematic diagram of a single periodic section of a travelling-wave parametric amplifier. Here $Z_{line}$ is the impedance of the transmission line section without periodic loading, $Z_{load}$ is the impedance of the transmission line section with periodic loading, $L_{per}$ is the length of periodic loading to achieve stopband at the third harmonic of the pump frequency $f_{pump}$, $L_{s}$ is the length of each loading element, and $L_{l}$ is the length of every third loading element.}
\end{figure}

Dispersion engineering in the form of periodic impedance loading is important to the operation of TWPAs \citenumns{Eom_2012, Chaudhuri_2015, Adamyan_2016, Shan_2016}. Impedance perturbations are introduced periodically along the transmission line (an example is shown in figure~\ref{fig:Periodic_Loading}) to create stop bands at higher pump harmonics, thereby reducing power loss due to shock formation \citenumns{Landauer_1960, Landauer_1960_2}. The calculation of $\gamma_{\alpha}$ requires more care due to the presence of these periodic impedance loadings. Here we apply Floquet analysis \citenumns{collin2001foundations} and divide the TWPA into identical periodic sections, as shown in figure~\ref{fig:Periodic_Loading}.

The potential difference and current before the n'th section are denoted by $V_{n}$ and $I_{n}$. Floquet analysis gives
\begin{align}
  \begin{bmatrix}
        V_{n} \\
        I_{n}
    \end{bmatrix} &=
  \begin{bmatrix}
      A & B\\
      C & D
  \end{bmatrix}    \begin{bmatrix}
        V_{n+1} \\
        I_{n+1}
    \end{bmatrix} \\
    &= e^{\gamma_{\alpha}l}\begin{bmatrix}
          V_{n+1} \\
          I_{n+1}
      \end{bmatrix} \, ,
\end{align}
where the matrix is the overall transfer matrix of the periodic section, $l$ is the total length of the  periodic section, and $\gamma_{\alpha}$ is the \textit{resultant} propagation constant for the $\alpha$-frequency mode. For the periodic structure shown in figure~\ref{fig:Periodic_Loading}, $l=3L_{per}$.

The transfer matrix can be computed by concatenating different components within each periodic section:
\begin{align}
  \begin{bmatrix}
      A & B\\
      C & D
  \end{bmatrix} &= \prod_{i}   \begin{bmatrix}
        A_i & B_i\\
        C_i & D_i
    \end{bmatrix}
  \\&= \prod_{i} \begin{bmatrix}
      \cosh{(\gamma_{\alpha,u} l_{i})} & Z_i\sinh{(\gamma_{\alpha,u} l_{i})}\\
      \sinh{(\gamma_{\alpha,u} l_{i})}/Z_i & \cosh{(\gamma_{\alpha,u} l_{i})}
  \end{bmatrix} \, ,
\end{align}
where $l_i$ is the length of the i-th component, $Z_i$ is the characteristic impedance of the i-th component, and $\gamma_{\alpha,u}$ is the propagation constant in the absence of periodic loading or non-linearity, and is given by
\begin{align}
  \gamma_{\alpha,u}^2 = -\omega_{\alpha}^2 LC+j(RC+GL)\omega_{\alpha}+RG\,.
\end{align}

\subsection{Extended Coupled-Mode Solver}

Previous coupled-mode solvers \citenumns{Chaudhuri_2015} assume that the signal  $A_s$ and idler $A_i$ modes are much weaker in magnitude than the pump mode $A_p$, and that the pump amplitude $|A_p|$ is constant along the transmission line. These assumptions need to be relaxed in order to analyse gain saturation. Here we only assume that all three primary modes (pump, signal, and idler) are significant and much higher in amplitude than the other harmonic modes, and that the pump amplitude is altered via coupling with signal and idler.

Applying the above approximations, the following system of differential equations can be obtained from equation~(\ref{eq:Combination}):
\begin{align}
\pdv{A_p}{z} =& \frac{K}{\gamma_p}\left[C_{p,p,-p}A_{p}A_{p}A_{-p}+C_{p,s,-s}A_{p}A_{s}A_{-s}\right. \notag \\&+\left.C_{p,i,-i}A_{p}A_{i}A_{-i}+C_{-p,s,-i}A_{-p}A_{s}A_{-i}\right] \\
\pdv{A_s}{z} =& \frac{K}{\gamma_s}\left[C_{s,p,-p}A_{s}A_{p}A_{-p}+C_{s,s,-s}A_{s}A_{s}A_{-s}\right. \notag \\&+\left.C_{s,i,-i}A_{s}A_{i}A_{-i}+C_{i,p,p}A_{i}A_{p}A_{p}\right]  \\
\pdv{A_i}{z} =& \frac{K}{\gamma_i}\left[C_{i,p,-p}A_{i}A_{p}A_{-p}+C_{i,s,-s}A_{i}A_{s}A_{-s}\right. \notag \\&+\left.C_{i,i,-i}A_{i}A_{i}A_{-i}+C_{s,-p,-p}A_{s}A_{-p}A_{-p}\right] \,,
\end{align}
where $K = {L_0C}/(8I_*^2)$. The coefficients are determined by substituting equations~(\ref{eq:CM_sum}-\ref{eq:CM_conj}) into equation (\ref{eq:Combination}), and are given by
\begin{align}
& C_{p,p,-p} = \omega_{p}^2 P_{p;p,p,-p} && C_{p,s,-s} = 2\omega_{p}^2 P_{p;p,s,-s} \notag \\
& C_{p,i,-i} = 2\omega_{p}^2 P_{p;p,i,-i} && C_{-p,s,-i} = 2\omega_{p}^2 P_{p;-p,s,-i} \notag \\
& C_{s,p,-p} = 2\omega_{s}^2 P_{s;s,p,-p} && C_{s,s,-s} = \omega_{s}^2 P_{s;s,s,-s} \notag \\
& C_{s,i,-i} = 2\omega_{s}^2 P_{s;s,i,-i}  && C_{i,p,p} = \omega_{s}^2 P_{s;i,p,p} \notag \\
& C_{i,p,-p} = 2\omega_{i}^2 P_{i;i,p,-p} && C_{i,s,-s} = 2\omega_{i}^2 P_{i;i,s,-s} \notag \\
& C_{i,i,-i} = \omega_{i}^2 P_{i;i,i,-i} && C_{s,-p,-p} = \omega_{i}^2 P_{i;s,-p,-p}\,,
\end{align}
where $P_{i;j,k,l}=\exp[(\gamma_{i}-\gamma_{j}-\gamma_{k}-\gamma_{l})z]$. Here we have taken the idler frequency to be $\omega_{i} = \omega_{s}-2\omega_{p}$.

The above system of coupled differential equations can be solved using a standard numerical package. In this study, solutions were obtained using the ODE45 algorithm of MATLAB. It is straightforward, if a little tedious, to incorporate higher harmonics into the above analysis scheme. We have not done so here because they do not produce significant qualitative differences, as shown in previous studies \citenumns{Chaudhuri_2015}.

\section{Simulations}

Values for $L$, $R$, $C$, and $G$ can be supplied by transmission-line theories such as \citenumns{Zhao_2018} for microstrip lines (MTL) or co-planar waveguides (CPW).

The series inductance $L_0$ and series resistance $R$ for a superconducting transmission line were calculated using the following routine:
\begin{enumerate}
  \item Compute the complex conductivity $\sigma=\sigma_1-j\sigma_2$ using \citenumns{Zhao_2017, MattisBardeen}.
  \item Compute the complex surface impedance $Z_s$ using \citenumns{Zhao_2017} - see Appendix regarding nuances in this calculation.
  \item Using transmission line theory \citenumns{Zhao_2018}, calculate the series impedance $Z$ from $Z_s$.
  \item The series inductance is given by $L_0=\operatorname{Im}(Z)/\omega$, and the series resistance is given by $R=\operatorname{Re}(Z)$.
\end{enumerate}

The shunt capacitance $C$ and shunt conductance $G$ were calculated using the following routine:
\begin{enumerate}
  \item Use tabulated data of $\operatorname{tan}(\delta)$ to obtain the complex dielectric constant
  \begin{align}
    \epsilon_{fm}=\epsilon'_{fm}-i\epsilon''_{fm}=\epsilon'_{fm}[1-j\operatorname{tan}(\delta)]\,,
  \end{align} where $\operatorname{tan}(\delta)$ is the loss tangent of the dielectric material \citenumns{lossTan_2008}.
  \item Using transmission line theory \citenumns{Zhao_2018}, calculate the shunt admittance $Y$ from $\epsilon_{fm}$.
  \item The shunt capacitance is given by $C=\operatorname{Im}(Y)/\omega$, and the shunt conductance is given by $G=\operatorname{Re}(Y)$.
\end{enumerate}

In the work reported here, we simulated a co-planar waveguide based on titanium nitride (TiN) with $L_0$ and $C$ as described in table~\ref{tab:table1}. Values of TiN properties were obtained from \citenumns{Gao_2014,Eom_2012}. Table~\ref{tab:table1} details the values of the demonstrative physical parameters used to compute the results in section \ref{sec:Results}:

\begin{table}[ht]
\begin{threeparttable}
\caption{\label{tab:table1}Table of parameters used in calculations.}
\begin{tabular}{b{0.40\linewidth} b{0.3\linewidth}}
\toprule
 & \textrm{TiN}\\
\colrule
$T_c$ (K) & 4 \tnote{a}  \\
$T$ (K) & 0.1 \tnote{a}  \\
$\rho_n$ ($\mathrm{\mu \Omega cm}$) & 100 \tnote{a}  \\
$I_*$ (mA) & 4 \tnote{a}  \\
$L_0$ ($\mathrm{\mu}$H) & 10 \tnote{a}\\
$C$ (nF) & 0.1 \tnote{a}\\
$f_{pump}$ (GHz) & 9.8\\
$L_l$ ($\mathrm{\mu}$m) & 10\\
$L_s$ ($\mathrm{\mu}$m) & 5\\
$L_{per}$ ($\mathrm{\mu}$m) & 530\\
$Z_{line}$ ($\mathrm{\Omega}$) & 315\\
$Z_{load}$ ($\mathrm{\Omega}$) & 150 \\
\toprule
\end{tabular}
\begin{tablenotes}[flushleft]
\RaggedRight
\footnotesize
\item[a] Values chosen to resemble TiN TWPA studied in \citenumns{Eom_2012}.
\end{tablenotes}
\end{threeparttable}
\end{table}

The range of $\tan{\delta}$ explored in this study is from $1\times10^{-5}$ to $1\times10^{-3}$. The range is chosen to reflect experimentally measured dielectric loss for various commonly used dielectric materials \citenumns{lossTan_2008} at low temperatures.

\section{Results}
\label{sec:Results}

\begin{figure}[!ht]
\includegraphics[width=8.6cm]{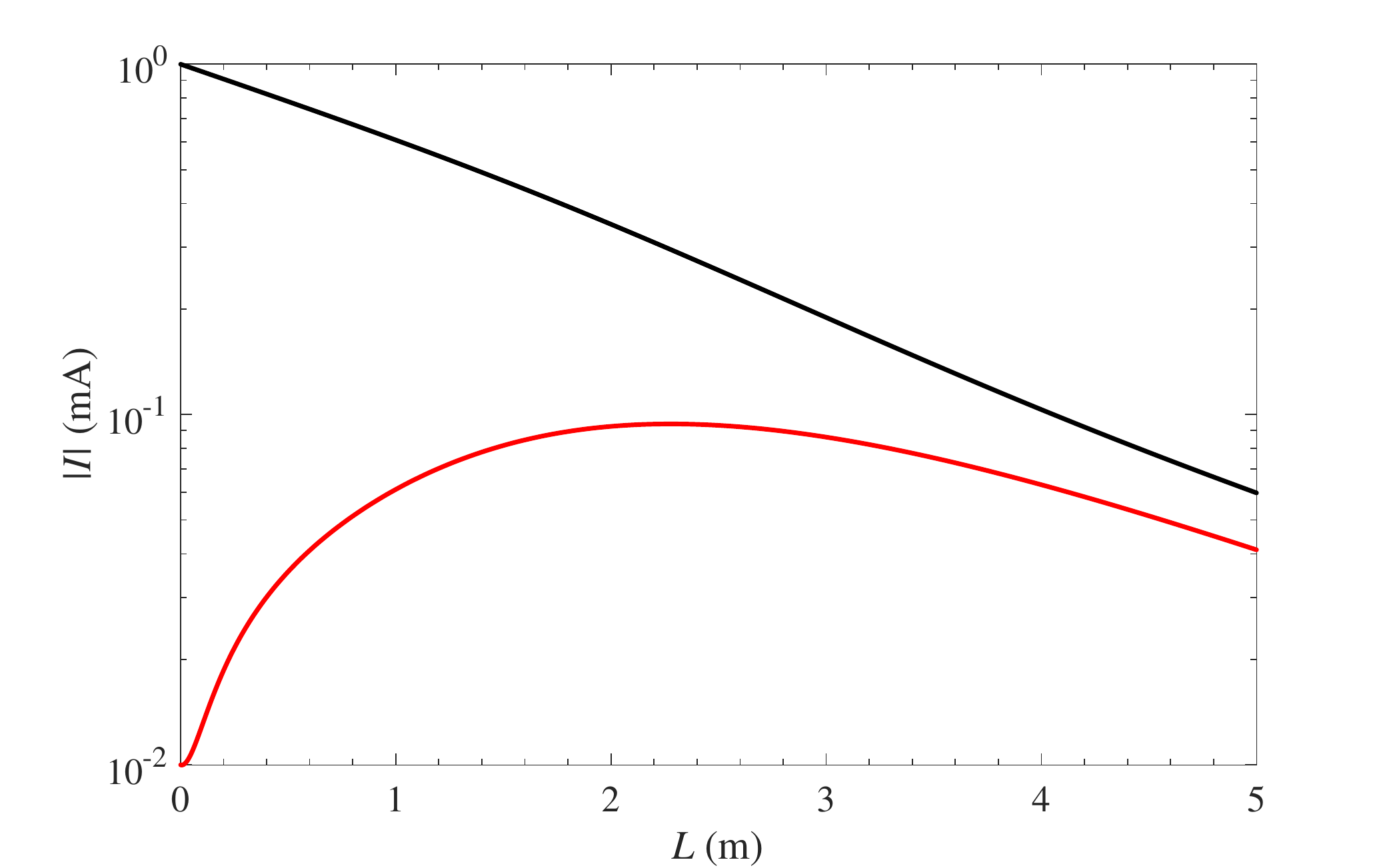}
\caption{\label{fig:Gain_Dist} Plot of amplitude of current modes $|I|$ against length $L$ for loss tangent $\tan{\delta}=5\times10^{-4}$, signal frequency $f=9\,\mathrm{GHz}$. a) Black line - pump mode; b) Red line - signal mode.}
\end{figure}

The behaviour of the output pump and signal current amplitude $|I|$  against length of the amplifier $L$ is shown in figure~\ref{fig:Gain_Dist} for loss tangent $\tan{\delta}=5\times10^{-4}$, signal frequency $f=9\,\mathrm{GHz}$. For short amplifiers, the energy transfer from the pump to the signal is relatively insignificant, and the signal increases with length similar to the predictions of previous studies \citenumns{Chaudhuri_2015}. For long amplifiers, the pump current mode is depleted both by energy transfer to the signal mode, as well as by loss in the transmission line. As a result, signal amplification cannot occur further and both the pump mode and the signal mode decay with length according to a linear lossy transmission line. A signal amplitude maximum can be found in between the two regimes.

\begin{figure}[!ht]
\includegraphics[width=8.6cm]{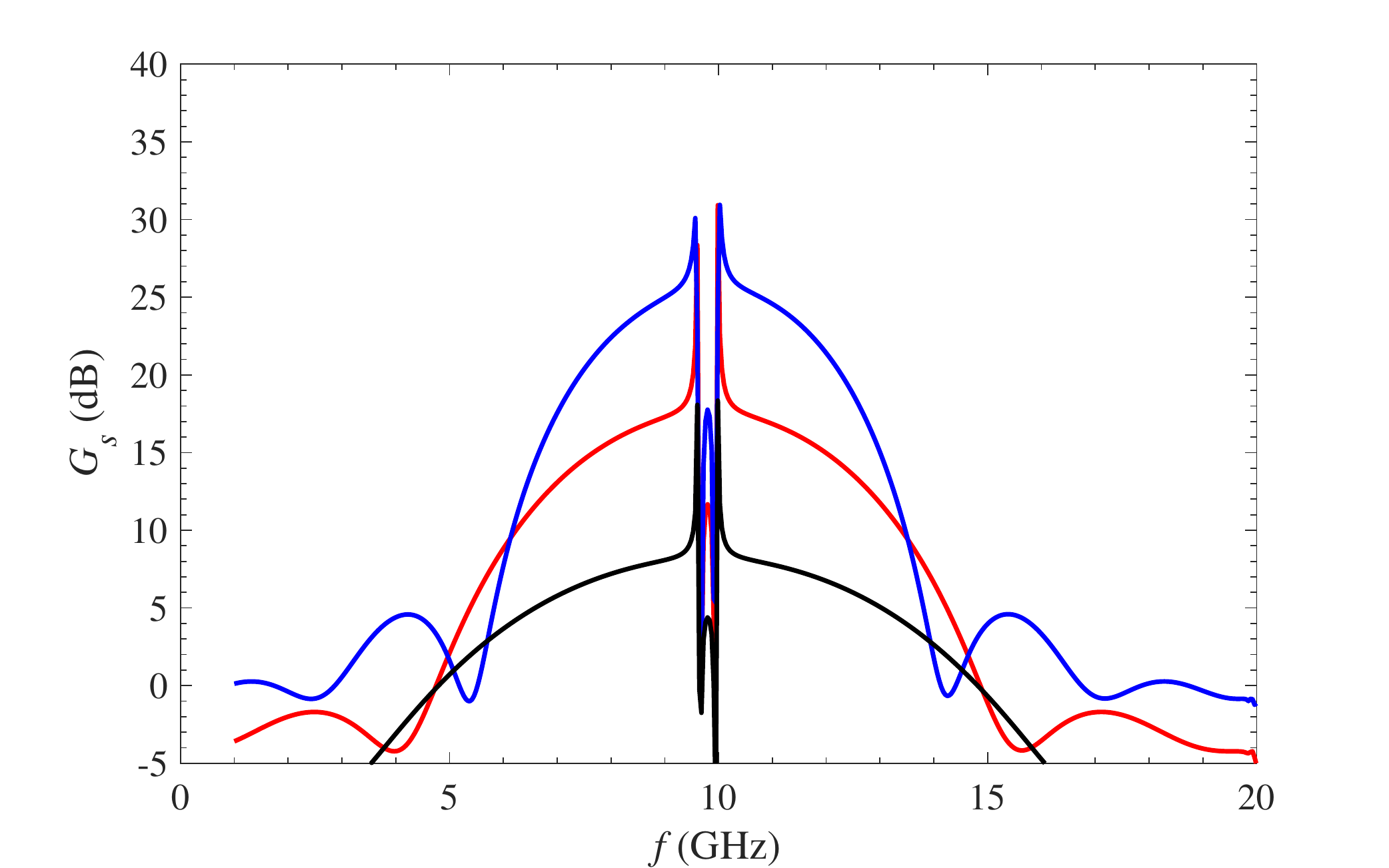}
\caption{\label{fig:Gain_Freq} Plot of gain of signal current mode $G_s$ against signal frequency $f$ for amplifier length $L=1\,\mathrm{m}$, and for various loss tangent values. a) Black line - $\tan{\delta}=1\times10^{-3}$; b) Red line - $\tan{\delta}=5\times10^{-4}$; c) Blue line - $\tan{\delta}=1\times10^{-4}$.}
\end{figure}

The behaviour of the signal current mode gain as a function of signal frequency is shown in figure~\ref{fig:Gain_Freq} for an amplifier length of $L=1\,\mathrm{m}$, and for loss tangents from $\tan{\delta}=1\times10^{-4}$ (blue line) to $\tan{\delta}=1\times10^{-3}$ (black line). The gain was calculated using
\begin{align}
 G_s = 20 \log_{10}\left(\frac{|I_{s,L}|}{|I_{s,0}|}\right)  \, ,
\end{align}
where $|I_{s,0}|$ is the input signal mode amplitude, and $|I_{s,L}|$ is the output signal mode amplitude. Figure~\ref{fig:Gain_Freq} is qualitatively similar to the experimentally measured gain profile as shown in figure~3 of \citenumns{Eom_2012}. We observe that as the loss tangent is decreased, the gain at frequencies close to the pump increases. The bandwidth at half-maximum gain decreases with decreased loss tangent. Similar effects can be obtained using a lossless TWPA model by increasing the input pump amplitude $|I_{p,0}|$. This is an intuitive result, which states that the effect of decreasing the loss tangent is similar to that of increasing the effective pump amplitude.

\begin{figure}[!ht]
\includegraphics[width=8.6cm]{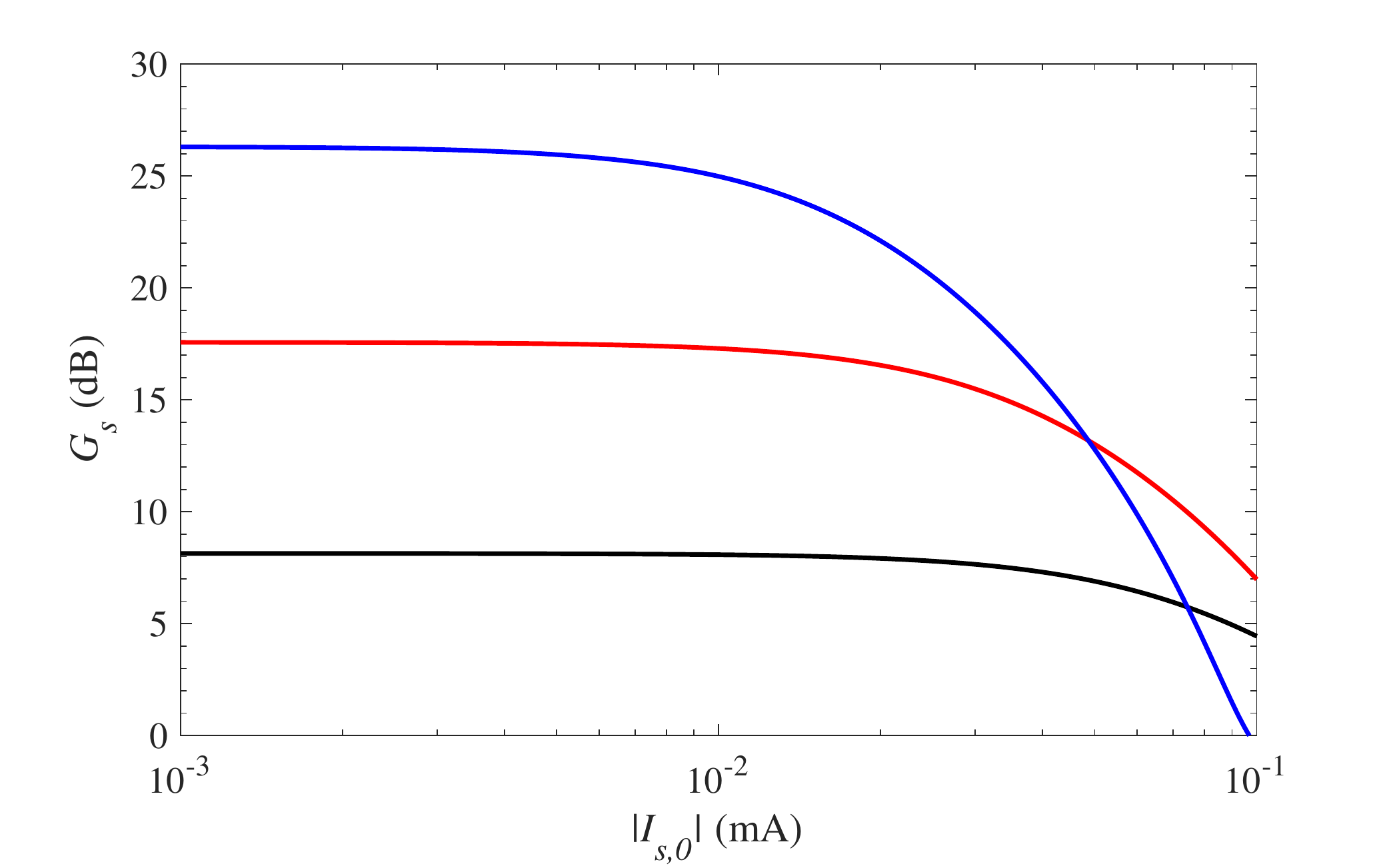}
\caption{\label{fig:Gain_Signal_TanDelta} Plot of gain of signal current mode $G_s$ against input amplitude of signal current mode $|I_{s,0}|$ for amplifier length $L=1\,\mathrm{m}$, signal frequency $f=9\,\mathrm{GHz}$. a) Black line - $\tan{\delta}=1\times10^{-3}$; b) Red line - $\tan{\delta}=5\times10^{-4}$; c) Blue line - $\tan{\delta}=1\times10^{-4}$.}
\end{figure}

\begin{figure}[!ht]
\includegraphics[width=8.6cm]{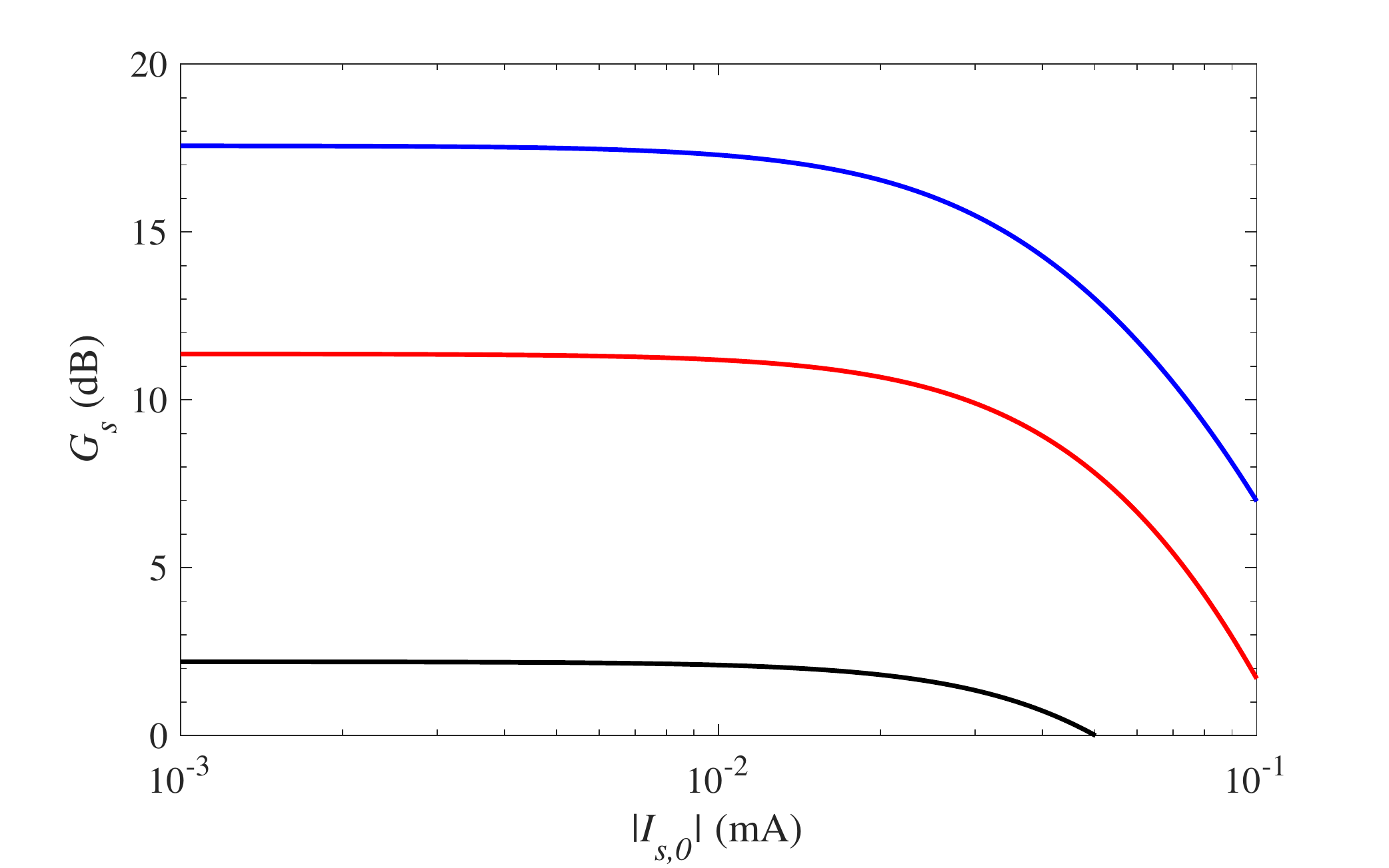}
\caption{\label{fig:Gain_Signal_Freq} Plot of gain of signal current modes $G_s$ against input amplitude of signal current mode $|I_{s,0}|$ for amplifier length $L=1\,\mathrm{m}$, $\tan{\delta}=5\times10^{-4}$. a) Black line - $f=5\,\mathrm{GHz}$; b) Red line - $f=6.5\,\mathrm{GHz}$; c) Blue line - $f=9\,\mathrm{GHz}$.}
\end{figure}

The behaviour of the signal mode gain $G_s$ as a function of input signal amplitude $|I_{s,0}|$ is shown in figure~\ref{fig:Gain_Signal_TanDelta} for loss tangent values from $\tan{\delta}=1\times10^{-4}$ (blue line) to $\tan{\delta}=1\times10^{-3}$ (black line), and in figure~\ref{fig:Gain_Signal_Freq} for signal frequencies from $f=9\,\mathrm{GHz}$ (blue line) to $f=5\,\mathrm{GHz}$ (black line). Figure~\ref{fig:Gain_Signal_Freq} is qualitatively similar to the experimentally observed gain saturation behaviour as shown in figure~S5 of \citenumns{Eom_2012}. Significant gain saturation occurs as the amplitude of the input signal is increased above some threshold. In fact, saturation occurs when the mixing modes that transfer energy out of the signal mode become significant compared with mixing modes that transfer energy from the pump mode to the signal mode. The rate of saturation is different for different frequencies and loss tangents.

\begin{figure}[!ht]
\includegraphics[width=8.6cm]{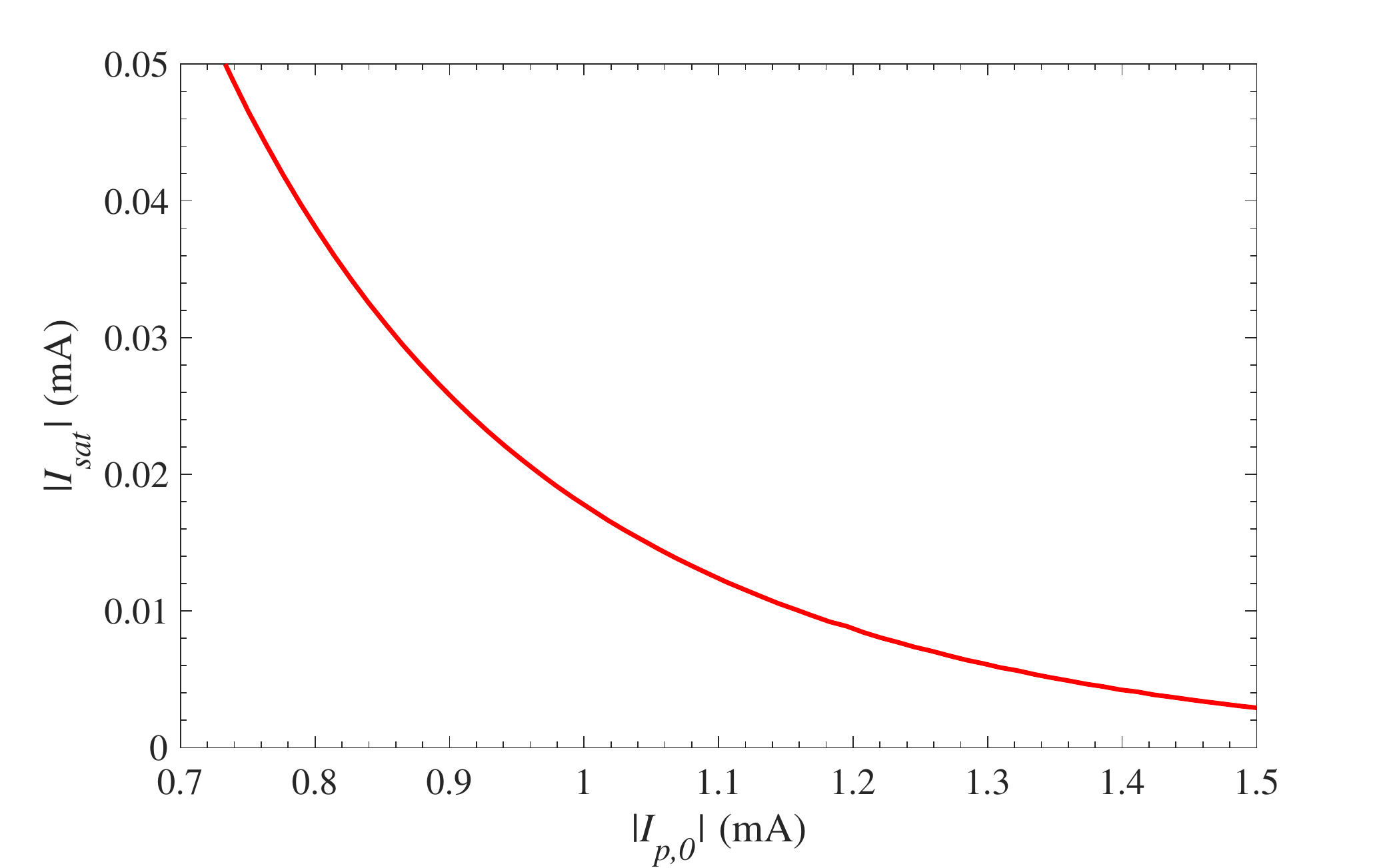}
\caption{\label{fig:Compression} Plot of 1dB compression point input signal amplitude $|I_{sat}|$ against pump amplitude $|I_{p,0}|$ for amplifier length $L=1\,\mathrm{m}$, signal frequency $f=9\,\mathrm{GHz}$, and $\tan{\delta}=5\times10^{-4}$.}
\end{figure}

Here we characterise gain saturation through the compression point, which is defined as the input signal amplitude $|I_{sat}|$ at which the signal gain decreases by 1dB with respect to the maximum. Figure~\ref{fig:Compression} shows $|I_{sat}|$ against input pump amplitude $|I_{p,0}|$ for an amplifier length of $L=1\,\mathrm{m}$, signal frequency $f=9\,\mathrm{GHz}$, and $\tan{\delta}=5\times10^{-4}$. As shown, $|I_{sat}|$ decreases with increasing $|I_{p,0}|$. Naively, one might expect $|I_{sat}|$ to increase with $|I_{p,0}|$ as the input signal-to-pump ratio decreases. However, the scale of current is set by $I_*$ which remains constant. Further, higher $|I_{p,0}|$ results in higher maximum gain, and hence the value of the gain that satisfies a 1dB compression is also higher. This result suggests a potential design trade-off between the signal gain (which increases with strong pump) and the saturation signal amplitude (which decreases with strong pump) when a strong pump is used.

\begin{figure}[!ht]
\includegraphics[width=8.6cm]{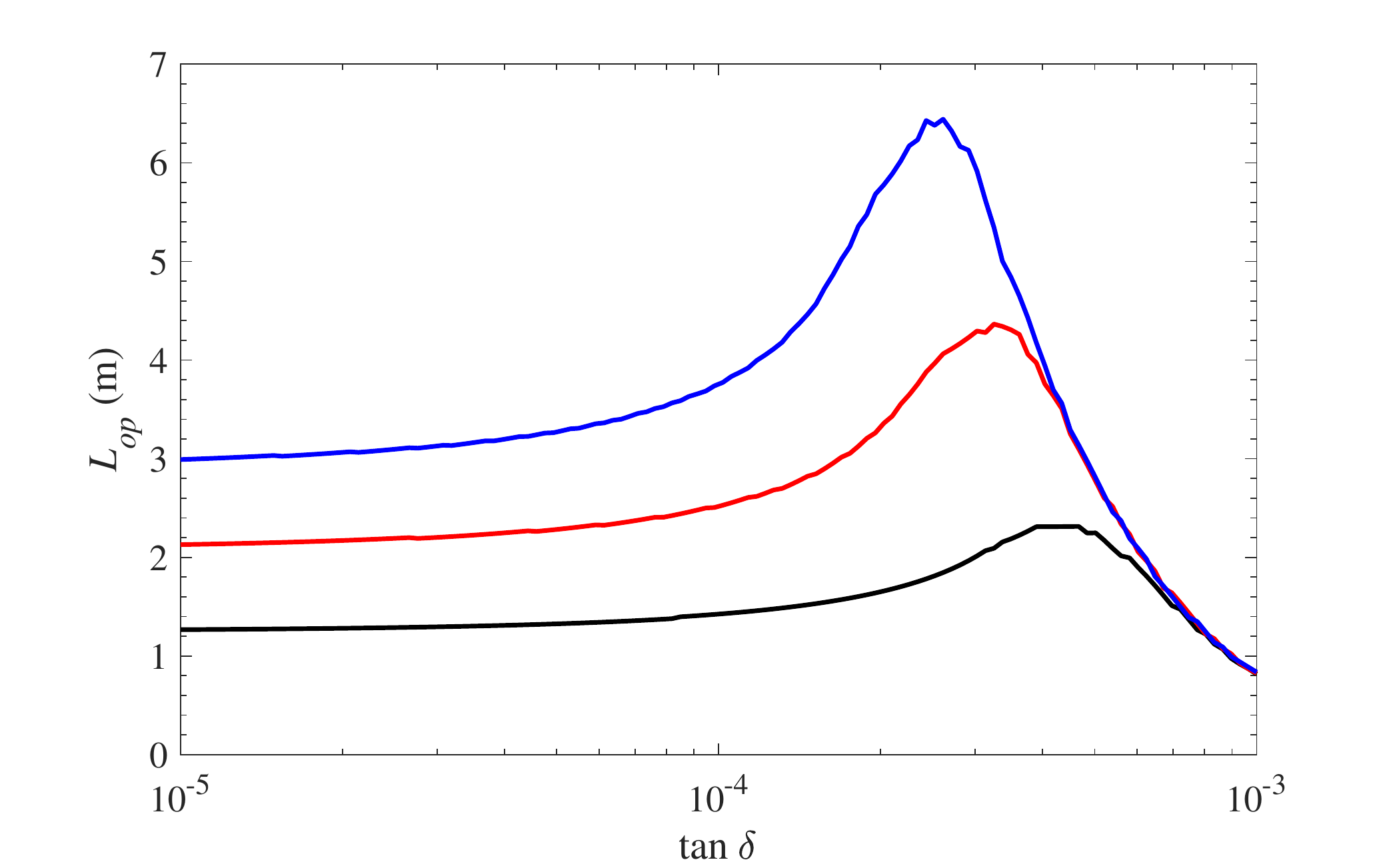}
\caption{\label{fig:Op_Length} Plot of optimum length $L_{op}$ against loss tangent $\tan{\delta}$ for signal frequency $f=9\,\mathrm{GHz}$, and pump amplitude $|I_{p,0}| = 1\,\mathrm{mA}$. a) Black line - signal amplitude $|I_{s,0}| = 10\,\mathrm{\mu A}$; b) Red line - signal amplitude $|I_{s,0}| = 1\,\mathrm{\mu A}$; c) Blue line - signal amplitude $|I_{s,0}| = 0.1\,\mathrm{\mu A}$.}
\end{figure}

\begin{figure}[!ht]
\includegraphics[width=8.6cm]{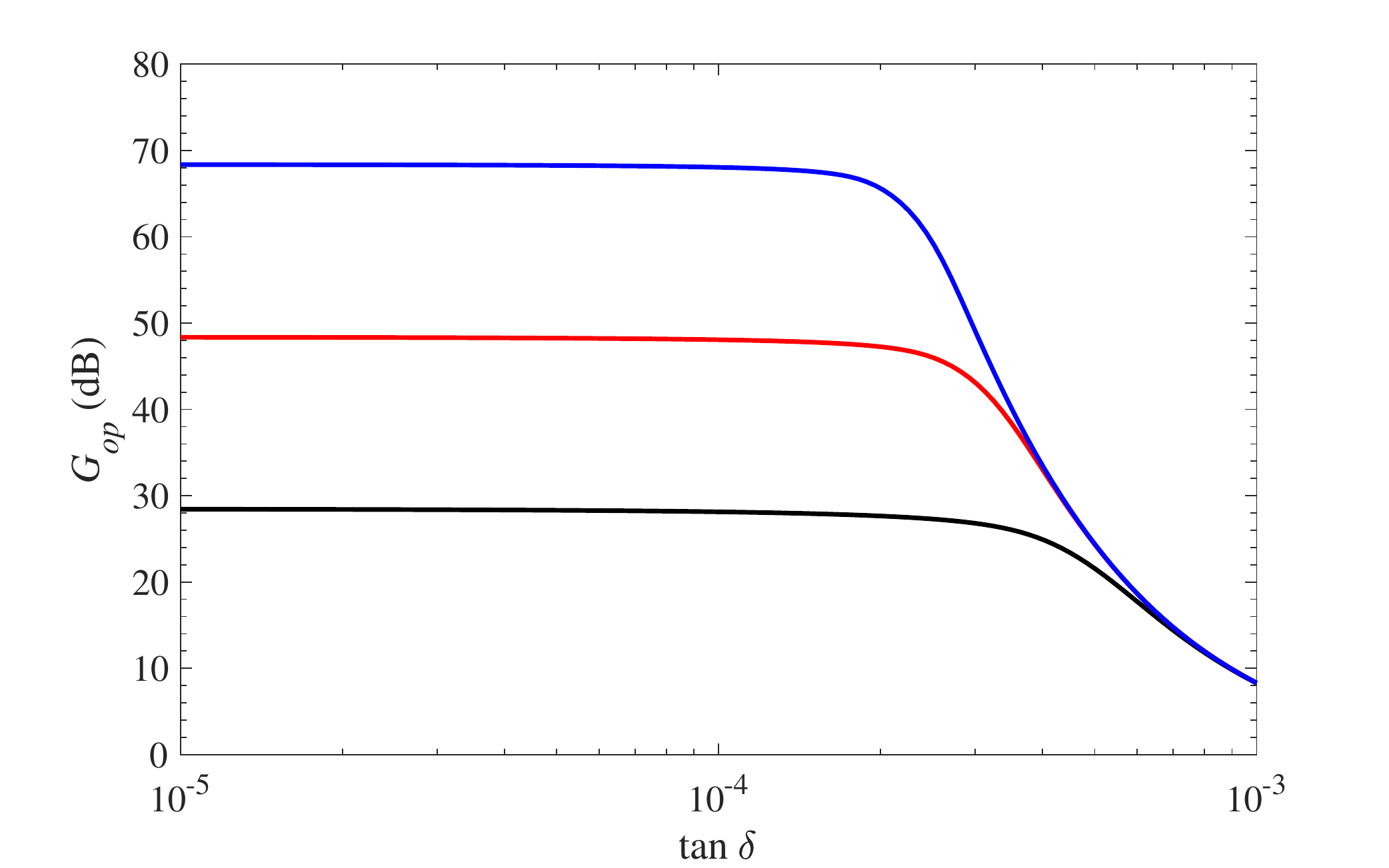}
\caption{\label{fig:Op_Gain} Plot of gain at optimum gain $G_{op}$ against loss tangent $\tan{\delta}$ for signal frequency $f=9\,\mathrm{GHz}$, and pump amplitude $|I_{p,0}| = 1\,\mathrm{mA}$. a) Black line - signal amplitude $|I_{s,0}| = 10\,\mathrm{\mu A}$; b) Red line - signal amplitude $|I_{s,0}| = 1\,\mathrm{\mu A}$; c) Blue line - signal amplitude $|I_{s,0}| = 0.1\,\mathrm{\mu A}$.}
\end{figure}

Figure~\ref{fig:Op_Length} shows the length of an amplifier $L_{op}$ for which maximum gain occurs, against loss tangent $\tan{\delta}$. Figure~\ref{fig:Op_Gain} shows the optimum gain $G_{op}$ (the value of gain maximum) against loss tangent $\tan{\delta}$. When $\tan{\delta}$ is small ($\tan{\delta}<10^{-4}$), gain maximum occurs when the amplitude of the amplified signal becomes comparable to that of the pump. In this regime greater values of $L_{op}$ are obtained for greater values of $\tan{\delta}$, and $G_{op}$ is approximately constant. When $\tan{\delta}$ is big ($\tan{\delta}>5\times10^{-4}$), gain maximum occurs when energy transfer from the pump is equal to the energy loss to the lossy transmission line. In this regime smaller values of $L_{op}$ and $G_{op}$ are obtained for greater values of $\tan{\delta}$. Knowing the optimum length and optimum gain is important when designing TWPAs. A length greater than the optimum should not be used to avoid the significant onset of transmission line loss, which is also likely to be associated with noise. If a gain greater than the optimum
is needed, this should be achieved by concatenating multiple TWPAs: to regenerate the pump amplitude. This could be achieved on a chip, for example, by using periodic directional couplers to inject a new pump signal. As $|I_{s,0}|$ decreases from $|I_{s,0}| = 10\,\mathrm{\mu A}$ (black line) to $|I_{s,0}| = 0.1\,\mathrm{\mu A}$ (blue line), when $\tan{\delta}$ is small ($\tan{\delta}<10^{-4}$), a greater amplifier length can be used before the signal amplitude becomes comparable to the pump amplitude. Hence $L_{op}$ increases when $|I_{s,0}|$ decreases. When $\tan{\delta}$ is big ($\tan{\delta}>5\times10^{-4}$), the signal amplitude does not become comparable to the pump amplitude regardless of the value of $|I_{s,0}|$. Hence the black, red, and blue lines in figure~\ref{fig:Op_Length} and figure~\ref{fig:Op_Gain} converge at high $\tan{\delta}$.

\begin{figure}[!ht]
\includegraphics[width=8.6cm]{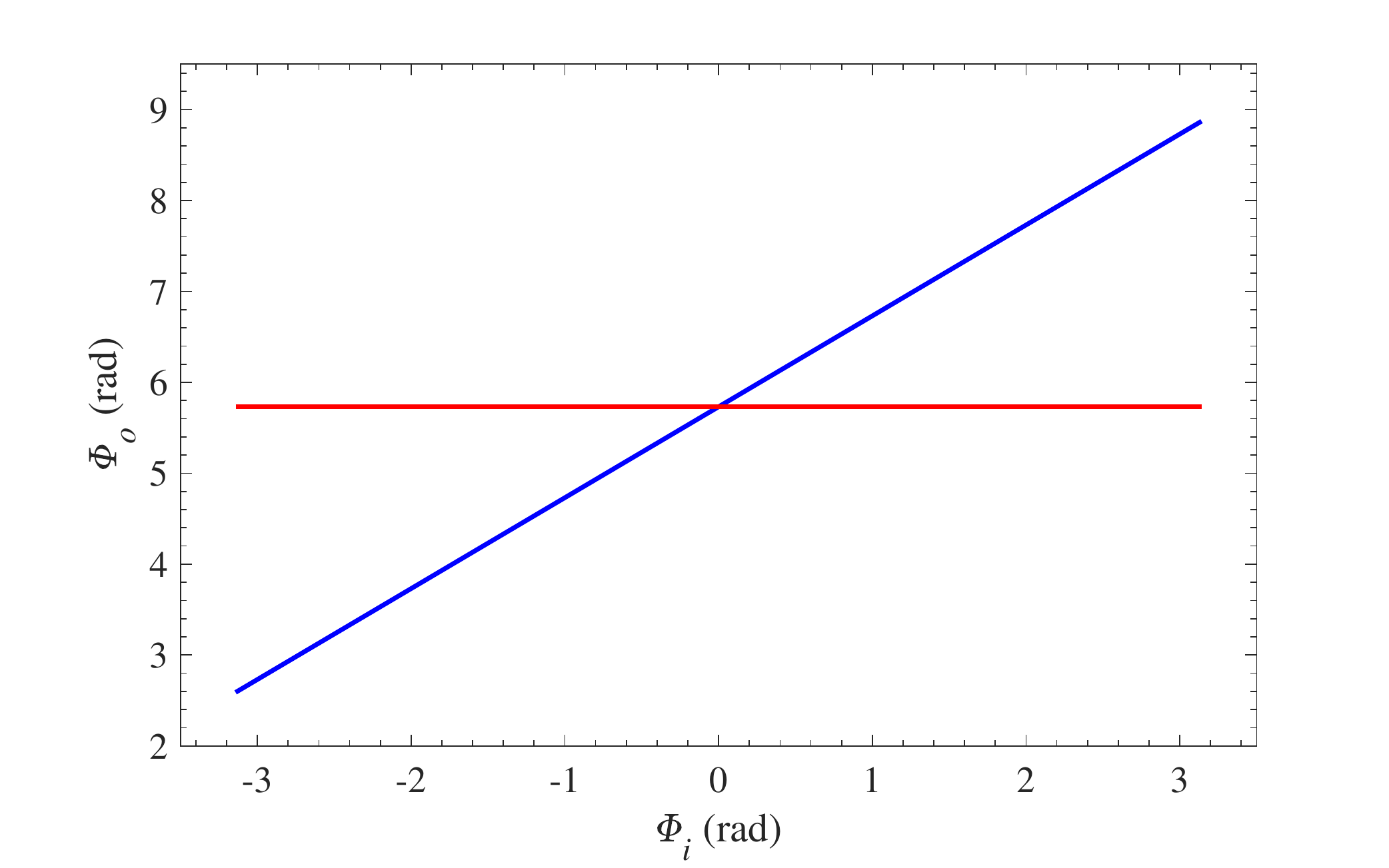}
\caption{\label{fig:Phase} a) Blue line - plot of output signal phase $\Phi_o$ against input signal phase $\Phi_i$ for amplifier length $L=1\,\mathrm{m}$, signal frequency $f=9\,\mathrm{GHz}$, and $\tan{\delta}=5\times10^{-4}$. b) Red line - plot of output signal phase $\Phi_o$ against input pump phase $\Phi_i$ for amplifier length $L=1\,\mathrm{m}$, signal frequency $f=9\,\mathrm{GHz}$, and $\tan{\delta}=5\times10^{-4}$.}
\end{figure}

The red line of figure~\ref{fig:Phase} shows output signal phase plotted as a function of the input signal phase. Within the analysis scheme, the amplifier is shown to be phase preserving. The blue line of figure~\ref{fig:Phase} shows the output signal phase as a function of the input pump phase. Here we observe that the phase of the amplified signal is insensitive to the phase stability of the pump, which has important consequences for achieving amplifiers operating with the quantum noise limit. This result agrees with predictions from previous analysis in \citenumns{Chaudhuri_2015}. The output signal phase stability against the input pump phase can be understood by noting that the signal mixing modes involve both the pump mode and its conjugate, and the input pump phase is cancelled out.

\begin{figure}[!ht]
\includegraphics[width=8.6cm]{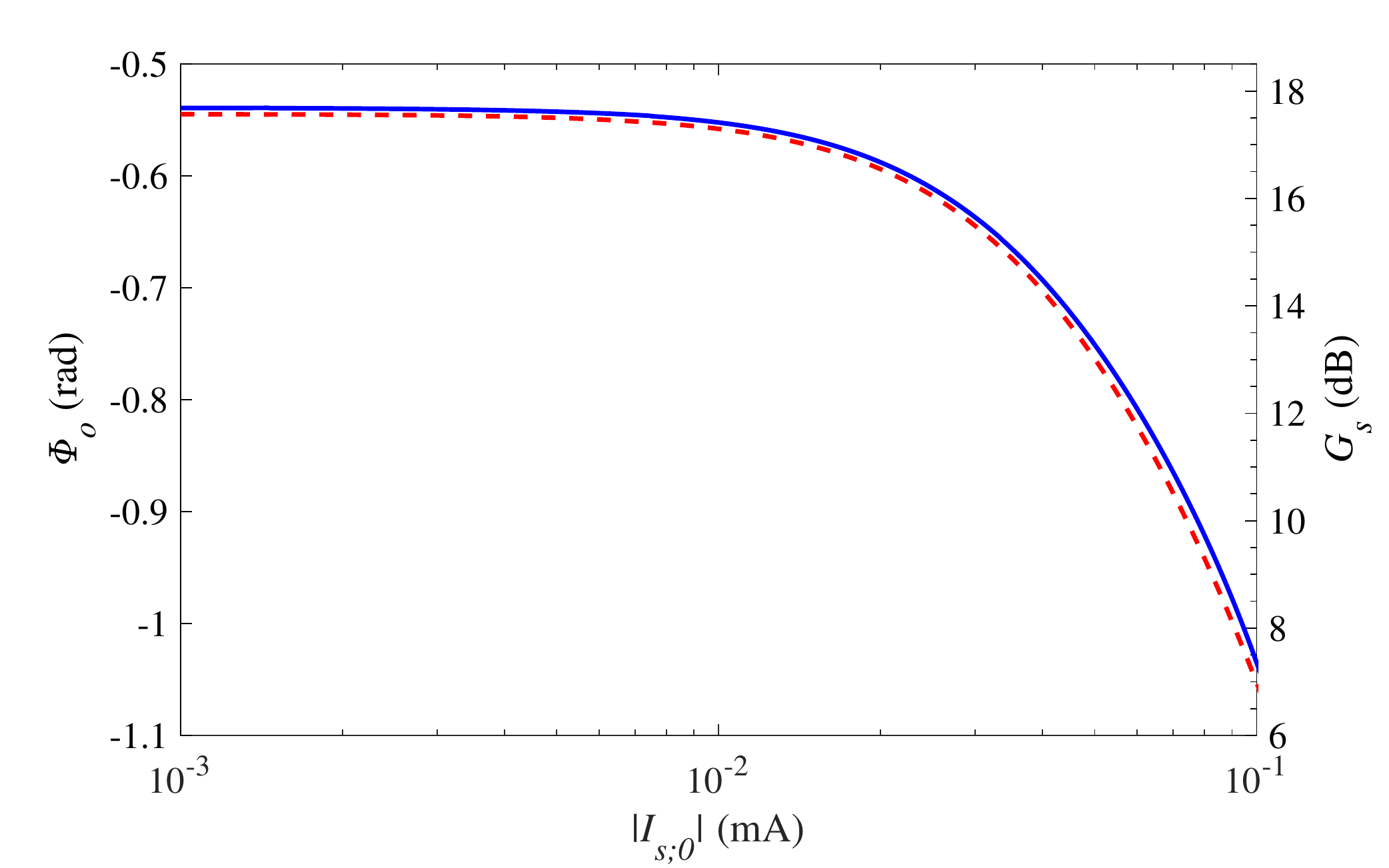}
\caption{\label{fig:Phase_Saturation} a) Blue, solid line, left y-axis - plot of output phase of signal current modes $\Phi_o$ against input amplitude of signal current mode $|I_{s,0}|$ for amplifier length $L=1\,\mathrm{m}$, signal frequency $f=9\,\mathrm{GHz}$, and $\tan{\delta}=5\times10^{-4}$. b) Red, dashed line, right y-axis - plot of gain of signal current modes $G_s$ against input amplitude of signal current mode $|I_{s,0}|$ for amplifier length $L=1\,\mathrm{m}$, signal frequency $f=9\,\mathrm{GHz}$, and $\tan{\delta}=5\times10^{-4}$.}
\end{figure}

Figure~\ref{fig:Phase_Saturation} shows the output signal phase $\Phi_o$ (blue, solid line) and gain $G_s$ (red, dashed line) against the input amplitude of signal current mode $|I_{s,0}|$. As seen, the change in the phase response is similar to the saturation in the amplitude response. Previous analyses (that do not account for gain saturation) have demonstrated $\log{G_s}\propto \Delta \Phi_s$ \citenumns{Adamyan_2016, Eom_2012, Chaudhuri_2015}, where $\Delta \Phi_s$ is the difference between the output signal phase and the input signal phase. Figure~\ref{fig:Phase_Saturation} shows that this relation is retained in the presence of large-signal gain saturation and phase shift. With regards to applications, this result demonstrates that the signal phase and gain change at the same rate,
and therefore the 1dB compression point metric can also be used for phase-sensitive applications, such as quantum non-demolition measurements \citenumns{Lupascu_2007}, qubit readout \citenumns{Blais_2004, Blais_2007}, and quantum information processing systems \citenumns{Charlene_2002,Bergeal2010,Silveri_2016}, as well as phase-insensitive applications.

\section{Discussion and Conclusion}

We have described a coupled-mode analysis of TWPAs that takes into account ohmic and dielectric loss in the transmission line system. It should be noted that even superconducting lines have ohmic loss at sufficiently high frequencies, and therefore this work has implications for understanding how high in frequency it should be possible to operate with a given material system. Our analysis does not assume that the pump mode is always much stronger than the signal and idler modes, and thus is able to replicate the gain saturation seen in TWPA experiments. Our model predicts the existence of a gain maximum against amplifier length. This puts a limit on the length, and therefore gain, of an optimum TWPA, indicating that higher gains should be achieved by concatenated designs that allow for pump regeneration. Our model also explains the onset of signal gain saturation in terms of energy loss of signal mode via frequency mixing. We have shown that the signal phase changes at a similar rate to the signal gain saturation as the amplitude of the signal is increased. Our model also predicts a trade-off between gain and saturation signal amplitude when a strong pump is used. Overall, these results are important in characterising the dynamics of a wide range of TWPAs, and allowing optimum designs, material systems, and drive conditions to be chosen.

\appendix
\section{Surface Impedance}

Complex surface impedance $Z_s=R_s+jX_s$ was calculated from complex conductivity $\sigma=\sigma_1-j\sigma_2$ using the transfer matrix method described in  \citenumns{Zhao_2017}. The transfer matrix method relies on a description of surface impedance that equates $Z_s$ seen by a normally incident plane wave with $Z_s$ seen by a wave parallel to the conductor \citenumns{Kerr1996}. It is important to distinguish between loss that occurs through total penetration and loss that occurs through dissipation. The importance of this effect can be demonstrated by calculating $Z_s$ at $T\rightarrow 0\,\mathrm{K}$, using a completely lossless $\sigma$. Despite the lack of dissipation in $\sigma$, a non-zero $R_s$ is obtained.

To the first order when $\sigma_2 \gg \sigma_1$, this problem can be addressed by stating that the transmission line surface resistance is given by
\begin{align}
  R_{s}=R_{s,TM}(\sigma_1-j\sigma_2)-R_{s,TM}(-j\sigma_2)\, , \label{eq:Z_s_difference}
\end{align}
where $R_{s,TM}$ is calculated using the original transfer matrix method.

It can be shown that for the case of homogeneous superconductor of thickness $t_s$, this calculation gives the same $R_s$ as the widely used formula \citenumns{Withington_1995}
\begin{align}
  R_s = \operatorname{Re}\left[\sqrt{\frac{j\omega\mu_0}{\sigma}}\operatorname{coth}\left(\sqrt{j\omega\mu_0\sigma}t_s\right)\right]\,, \label{eq:Z_s_original}
\end{align}
where $\omega$ is the angular frequency of the wave on transmission line, and $\mu_0$ is the vacuum permeability. This agreement lends confidence to the treatment detailed in this section. Equation~(\ref{eq:Z_s_original}) is derived in \citenumns{Kerr1996} by assuming that the impedance of free space $Z_{\eta}\gg\left|\sqrt{{\omega\mu_0}/{\sigma}}\right|$. Equation~(\ref{eq:Z_s_difference}) removes the need of using this assumption by addressing directly the contributions to $R_s$, and can be used to calculate surface impedances of superconducting multilayers as part of the transfer matrix method.

\bibliographystyle{h-physrev}
\bibliography{library}
\end{document}